\documentclass{emulateapj}
\usepackage{amsmath}
\usepackage{epsfig}
\usepackage[dvips]{color}
\def\ngrb{\dot{N}_{\rm{G}}}
\def\rhogrb{\rho_{\rm{G}}}
\def\rgrb{G}
\def\rhoe{\rho_{\rm emp}}
\def\grbII{G_{\rm II}}
\def\RII{R_{\rm II}}
\def\MII{M_{\rm II}}
\def\sII{\sigma_{\rm II}}
\def\gII{\gamma_{\rm II}}
\def\imfII{{\cal I}_{\rm II}}
\def\grbIII{G_{\rm III}}
\def\RIII{R_{\rm III}}
\def\MIII{M_{\rm III}}
\def\sIII{\sigma_{\rm III}}
\def\gIII{\gamma_{\rm III}}
\def\imfIII{{\cal I}_{\rm III}}
\def\Eiso{E_{\rm iso}}
\def\ris{r_{\rm is}}
\def\Ejet{E_{\rm{jet}}}
\def\Gjet{\Gamma_{\rm jet}}
\def\xacc{\xi_{\rm acc}}
\def\xBjet{\xi_{B}^{\rm{jet}}}
\def\UBjet{U_{B}^{\rm jet}}
\def\UBh{U_{B}^{\rm h}}
\def\Ug{U_\gamma^{\rm jet}}
\def\Ugh{U_\gamma^{\rm h}}
\def\Up{U_p^{\rm jet}}
\def\Tjet{T_{\rm jet}}
\def\Th{T_{\rm h}}
\def\xBjh{\xi_B^{\rm h}}
\def\rh{r_{\rm h}}
\def\tacc{t_{\rm acc}}
\def\Gh{\Gamma_{\rm h}}
\def\Gprime{\Gamma^{\prime}}
\def\npj{n_p^{\rm jet}}
\def\fpj{f_p^{\rm jet}}
\def\nph{n_p^{\rm h}}
\newcommand{\be}{\begin{equation}}
\newcommand{\ee}{\end{equation}}
\newcommand{\bea}{\begin{eqnarray}}
\newcommand{\eea}{\end{eqnarray}}
\newcommand{\bw}{\begin{widetext}}
\newcommand{\ew}{\end{widetext}}
\def\d{{\rm d}}
\def\alt{\raise0.3ex\hbox{$\;<$\kern-0.75em\raise-1.1ex\hbox{$\sim\;$}}}
\def\agt{\raise0.3ex\hbox{$\;>$\kern-0.75em\raise-1.1ex\hbox{$\sim\;$}}}

\begin{document}
\title{High Energy neutrino signals from the Epoch of Reionization}
\author{F. Iocco$^{1,2}$, K. Murase$^{3}$, S. Nagataki$^{2,3}$, P.~D.~Serpico$^{4}$}
\affiliation{$^1$Universit\`a di Napoli ``Federico II",
Dip. Scienze Fisiche, via Cintia, 80126 Napoli, Italy\\
$^2$Kavli Institute for Particle Astrophysics and Cosmology
PO Box 20450, Stanford, CA 94309, USA\\
$^3$YITP, Kyoto University, Kyoto, Oiwake-cho, Kitashirakawa,
Sakyo-ku, Kyoto, 606-8502, Japan\\
$^4$Center for Particle Astrophysics, Fermi National Accelerator
Laboratory, Batavia, IL 60510-0500 USA}

\begin{abstract}
In this paper we perform a new estimate of the high energy neutrinos
expected from GRBs associated with the first generation of stars in
light of new models and constraints on the epoch of reionization and
a more detailed evaluation of the neutrino emission yields. We also
compare the diffuse high energy neutrino background from Population
III stars with the one from ``ordinary stars" (Population II), as
estimated consistently within the same cosmological and
astrophysical assumptions. In disagreement with previous literature,
we find that high energy neutrinos from Population III stars will
not be observable at current or near-future neutrino telescopes,
falling below both the sensitivity of a km$^3$ telescope and the
atmospheric neutrino background, also under the most optimistic
predictions for the GRB rate. This rules them out as a viable
diagnostic tool for these still elusive metal-free stars.
\end{abstract}


%%%%%%%%%%%%%%%%%%%%%%%%%%%%%%%%%%%%%%%%
\section{Introduction}\label{sec:intro}
%%%%%%%%%%%%%%%%%%%%%%%%%%%%%%%%%%%%%%%%
The first generation of stars (Population III or PopIII stars) born
after the collapse of the very first structures in the universe is
puzzling the astrophysical community since long time. The vanishing
metallicity of the universe at the epoch of their formation
\cite{Iocco:2007km} is supposed to give them peculiar properties,
all arising from their very high characteristic mass due to the
peculiar cooling of the cloud \cite{Abel:2001pr}. In fact, stars of
${\cal O}$(100)$\,M_{\odot}$ have very short lifetimes and either
explode as pair instability Super Novae (PISNe) or directly collapse
to black holes \cite{Heger:03}. They also contribute to the
metallicity enrichment of the early universe, shaping the transition
to the second generation of stars. Moreover, the very high
temperature of these objects makes them efficient engines for the
production of Lyman-Werner ultraviolet photons, thus initiating the
cosmic reionization process. Unfortunately, all of the traces PopIII
stars leave behind are challenging to observe and so far no
unambiguous detection has been recognized, although the detection of
anisotropies in the infrared background consistent with the
existence of PopIII has been claimed \cite{Kashlinsky:2005di}.

Despite the difficulty of detecting neutrinos, they may be
potentially interesting messengers of the high-redshift universe
since, differently from gamma rays, neutrinos travel unimpeded over
cosmological distances. Also, the peculiar initial mass function
(IMF) of PopIII stars should increase their Super Nova (SN) rate
with respect to later stellar populations. The peculiarity of PopIII
stars may well produce high energy neutrinos above the expectations
for PopII stars, thus making them an intriguing target for neutrino
telescopes. A fairly general expectation is that PopIII may emit
large amounts of neutrinos, both thermally produced at the time of
the collapse \cite{Iocco:05} and non-thermally during a Gamma-Ray
Burst (GRB) phase associated with the explosion
\cite{Schneider:2002sy}. GRBs are one of the candidate hadronic
accelerators, and any proton accelerator is a potential emitter of
high energy neutrinos, produced via hadronic ($pp$) or
photo-hadronic ($p\gamma$) reactions in the surrounding medium.
Although the details of this whole picture are far from being
established and despite the large uncertainties in the models, the
idea of using neutrinos to probe PopIII stars seems promising, and
surely deserves further study.

In this paper we perform a new estimate of the high energy neutrinos
expected from GRBs associated with the first generation of stars in
light of new models and constraints on the epoch of reionization
(EoR) and a more detailed evaluation of the neutrino emission
yields.  The goal of this paper is two-fold: $(i)$ to compare the
diffuse high energy neutrino background from PopIII with the one
from ``ordinary" (PopII) stars that contribute with PopIII to the
reionization at $7\alt z\alt 12$; $(ii)$ to discuss the chances of
detection on the light of the performances expected for current or
future neutrino telescopes as a function of the astrophysical input.
Indeed both conditions need to be fulfilled to establish if there is
any realistic chance to discriminate the high energy neutrino
emission from PopIII. We anticipate that we find that high energy
neutrinos from Population III stars will not be observable at
current or near future neutrino telescopes, falling below both
IceCube sensitivity and atmospheric neutrino background under the
most optimistic assumptions for the GRB rate. The disagreement with
previous literature is mostly due to unrealistically high PopIII GRB
rates previously considered, following e.g. from values of the
ionization efficiency nowadays considered too extreme based on
self-consistent reionization scenarios.

The plan of this paper is the following: after introducing the basic
formalism in Sec. \ref{sec:formalism}, we shall preliminarily
discuss what a semi-empirical estimate of the GRB rate at high
redshift can tell us (Sec. \ref{sec:Obser}). Then, we devote Sec.
\ref{sec:TheorModel} to describe the models used to perform
theoretical estimates of the neutrino fluxes. We shall pay
particular attention to compare estimates derived consistently
within the same cosmological assumptions. In Sec. \ref{sec:PrelRes}
we present our results and compare with the chances of detection,
and finally in Sec. \ref{conclusion} we conclude. In Appendix
\ref{modelsummary} we report some detail of the GRB model used to
compute the neutrino yields.

%%%%%%%%%%%%%%%%%%%%%%%%%%%%%%%%%%%%%%%%
\section{Basic formalism}\label{sec:formalism}
%%%%%%%%%%%%%%%%%%%%%%%%%%%%%%%%%%%%%%%%
In this Section we establish the formalism we will use throughout
the paper to estimate the diffuse flux of neutrinos emitted by GRBs.
The integrated signal observed today at energy $E_{\nu}$ is
\bea E_{\nu}^2 \Phi_{\nu}  \,\, [{\rm GeV \, cm}^{-2}\, {\rm
s}^{-1}\, {\rm sr}^{-1}] \equiv E_{\nu}^2\frac{\d F_{\nu}}{\d
E_{\nu}}=\nonumber \\E_{\nu}^2 \int \d z \frac{\d \ngrb(z)}{\d z\d
\Omega} \frac{\d N_{\nu}^{\rm{iso}}}{\d E_{\nu} \d A}(E_{\nu},z),
\eea
where $\d \ngrb/\d z\d \Omega$ is the differential rate of GRBs
which beam towards us per unit solid angle, and $\d
N_{\nu}^{\rm{iso}}/\d E_{\nu} \d A$ is the average flux emitted by a
single source at energy $E_{\nu}(1+z)$. The function $\d \ngrb/\d
z\d \Omega$ can be written as
\be
 \frac{\d \ngrb}{\d z\d \Omega}(z)=\frac{1}{4 \pi}\frac{\rhogrb(z)}{(1+z)}\frac{\d V}{\d z}
=\frac{b}{4\pi}\frac{\rgrb(z)}{(1+z)}\frac{\d V}{\d z} , \ee
where $\rhogrb(z)$ is the GRB rate in a comoving volume and $\d V$
is the comoving volume element, such that
\be \label{vol} \frac{\d V}{\d z} = 4 \pi \frac{c}{H(z)} \,
r^2(z)\,. \ee
We have introduced the comoving distance $r(z)$ defined as
\be \label{comdist} r(z)=\int_0^z \frac{c}{H(w)}\,\d w\,. \ee
The Hubble function $H(z)$ writes in terms of the fractions of the
critical energy density in matter $\Omega_M$ and cosmological
constant $\Omega_\Lambda$ as
\begin{equation}
H(z) = H_0 \sqrt{\Omega_M(1+z)^3 + \Omega_k(1+z)^2
+\Omega_\Lambda}\,,
\end{equation}
$H_0$ being the inverse Hubble distance today $H_0^{-1}\simeq
3\,h^{-1} $Gpc and $\Omega_k=1-\Omega_M-\Omega_\Lambda$. We have
also denoted by $b=\langle 1-\cos \theta_{\rm jet}\rangle$ the
averaged beaming factor of a jet of opening angle $\theta_{\rm
jet}$, so that $\rgrb(z)$ is the beaming-corrected overall GRB rate.
The quantity $b$ also fixes the ratio between the physical energy
$\Ejet$ and the one released by the GRB if that was isotropic,
$b/4\pi=\Ejet/\Eiso$. If we introduce the function $J_{\nu}$ of the
energy at the source $E_{\nu}^{\prime}$
\be J_{\nu}[E_{\nu}^{\prime}] \equiv {E_{\nu}^{\prime}}^2 \frac{\d
N_{\nu}^{\rm{iso}}}{\d E_{\nu}^{\prime}}\,, \ee
the average energy flux, $E_{\nu}^2 (\d N_{\nu}^{\rm{iso}}/\d E_{\nu} \d A)$
(units GeV cm$^{-2}$), observed at a given energy $E_{\nu}$ and
emitted by a source at redshift $z$ can be expressed as
\be E_{\nu}^2 \frac{\d N_{\nu}^{\rm{iso}}}{\d E_{\nu} \d A}(E_{\nu},z) =
\frac{1}{1+z} \frac{J_{\nu}[E_{\nu} (1+z)]}{4 \pi r^2(z)}\,.
\label{eq:single} \ee
In the previous expression, $E_{\nu}(1+z)$ is the emission energy,
$J_{\nu}[E_{\nu} (1+z)]/ (1+z)$ takes into account the redshifted
energy spectrum, and possible boosting factors are already included
in $J_\nu$.

Putting the two pieces together, we finally get
\be E_{\nu}^2\Phi_{\nu} = \frac{c\, b}{4 \pi} \int \d z
\frac{J_{\nu}[E(1+z)] \, \rgrb (z)}{(1+z)^2 H(z)}.
\label{eq:summary} \ee
It remains to estimate $\rgrb(z)$ and $J_\nu(E)$, a problem which we
shall address in the following.

%%%%%%%%%%%%%%%%%%%%%%%%%%%%%%%%%%%%%%%%%
\section{An empirical estimate of GRB rate}\label{sec:Obser}
%%%%%%%%%%%%%%%%%%%%%%%%%%%%%%%%%%%%%%%%%
Let us start by using an empirical approach: the GRB rate is derived
from observations and extrapolated at high redshifts, as in
\cite{Yonetoku:2003gi} and \cite{FenRamRui}. This approach {\it
assumes} the $E_{p}$--luminosity relation, \cite{Amati:02}, to infer
the redshift of a GRB whose host is at unknown redshift. Note that
in order to extrapolate the $E_{p}$--luminosity relation to high
redshifts, one implicitly assumes that although GRBs at high
redshift might belong to different stellar populations with
different star formation rates (SFRs) and IMFs, they constitute
``standard candles" carrying no memory of the population of the
progenitors they belong to.

Aware of this {\it caveat}, the most natural hint of a transition
PopIII-PopII would then be a break of the power-law dependence of
the GRB formation history on the redshift; indeed, one should expect
in general that the conditions for the GRB appearance will occur at
a different rate in stellar populations with different
characteristics. This is manifestly not the case of the GRB rate
derived empirically from observations in \cite{Yonetoku:2003gi} and
\cite{FenRamRui}: no discontinuity appears at 6$<z<$12, where a
PopIII-PopII transition would suggest to see it. In particular,
these results indicate that the GRB formation rate always increases
toward $z \sim 12$, in a way parameterized by:
\begin{eqnarray}
\label{eq:l-theta}\rhoe (z) =\rhoe(0) \times\left\{
\begin{array}{ll}
    (1+z)^\alpha\,,
        &\, z <1 \\
    2^{\alpha-\beta} (1+z)^\beta\,,
        &\,  1<z\alt 12, \\
\end{array}
\right.
\end{eqnarray}
where $\alpha=6.0\pm1.4$, $\beta=0.4\pm 0.2$ and
\be \rhoe(0)=2.3\times10^{-14} {\rm s}^{-1} {\rm Mpc}^{-3}\,, \ee
all the quantities being fixed according to \cite{Murakami:05},
where for the normalization we use their intermediate estimate for
the luminosity correction ($k=0.5$). Note that if we assume a simple
top-hat function for $J_\nu$, from Eq. (\ref{eq:summary}) and the
fit for $\rhoe(z)$ it is easy to check that neutrinos emitted
at high redshift have at most a subleading contribution to
the total flux. Qualitatively, this is consistent with what found
e.g. in \cite{Murase:2005hy}: most of the GRB neutrino signal is not
sensitive to the GRB rate at high redshift.

This general result might be interpreted as (weak) empirical
evidence for the dominance of PopII progenitors in the overall GRB
population at any redshift. However, this conclusion only holds
under two assumptions: $(i)$ the validity of the $E_{p}$--luminosity
relation up to high redshifts; since this is an empirical relation
and the engine of GRB is far from being understood, there is no
guarantee that this is the case; $(ii)$ the absence of a significant
fraction of {\it choked} GRBs among PopIII stars. It has been argued
that due to the different structure and properties of PopIII stars
the jet in a collapsar may be unable in most cases to punch through
the stellar envelope \cite{MacFadyen:1999mk}, thus choking the burst
in gamma rays, although still producing large neutrino yields. This
was also the working hypothesis of \cite{Schneider:2002sy}. These
arguments suggest that a more realistic estimate of the neutrino
flux from PopIII GRB may require some degree of modeling, which we
are going to address in the next section. However, we can anticipate
that the empirical GRB rate obtained in this section is roughly in
agreement with the one we obtain from the CF05 model we shall
introduce in the following section.

%%%%%%%%%%%%%%%%%%%%%%%%%%%%%%%%%%%%%%%%%
\section{Theoretical estimates of the neutrino flux}\label{sec:TheorModel}
%%%%%%%%%%%%%%%%%%%%%%%%%%%%%%%%%%%%%%%%%
Let us now estimate the GRB rate at high redshifts from current
theoretical models on the stellar populations at high redshift,
under the assumption that the GRB rate tracks the SFR. Besides the
SFR for the different stellar populations, it is clearly mandatory
to know their IMFs in order to evaluate the fraction of stars that
are likely to end their lives as collapsars, believed to give rise
to a GRB. In this section we present the neutrino fluxes expected
from the two different populations of stars under different EoR
models available in literature. The three models of reionization we
consider are summarized in Table \ref{EoRModels}: the ``fiducial"
model presented in \cite{Choudhury:2004vs}, figure 1 at page 586
(hereafter CF0$5_a$); the one in figure 4, page 590 of the same
paper (hereafter CF0$5_b$); the model in \cite{Choudhury:06}
(hereafter CF06), which takes into account the new cosmological data
on reionization from the third-year data release of the WMAP team
\cite{WMAP3}. These models are used in the following to derive the
star formation rates needed for our estimates. All the CF-models
share the same physical assumptions: three radiation sources are
taken into account during EoR: Quasi Stellar Objects (QSOs), PopII,
and PopIII stars which contribute with different characteristics to
the radiative and chemical enrichment of the IGM, and to the
consequent feedback. Among the many parameters of the models, a key
role is played by the fraction of escaping photons per halo
(directly related to the ionizing efficiency), which must be deduced
from the knowledge of the source and halo characteristics of PopII
and PopIII stars; the luminosity function of QSOs is calculated
according to standard assumptions. Both the CF05 models assume that
all PopIII stars have a mass of 300M$_{\odot}$ while PopII follow a
Salpeter IMF, $S(m)$. It is also worth mentioning that the
transition between PopIII and PopII stars, regulated by the chemical
feedback, is considered to be instantaneous. The only difference
between CF05$_a$ ad CF05$_b$ is the ionizing efficiency of PopIII,
which in the latter model is assumed to be 2/7 of the former one.
This drastically changes the normalization and shape of the PopIII
SFR. With respect to the CF05 models, in CF06 the stellar formation
in minihaloes is suppressed, and a self--consistent calculation is
made of the chemical enrichment process, which takes longer and
results in a non--instantaneous transition between PopII and PopIII.
For our purposes, however, the crucial difference is that CF06
assumes a Salpeter IMF for the PopIII, too. The authors implement
this choice in order to match the NICMOS high redshift source
counts. We address the reader to the original papers for more
details. Also note that none of the models provides the low-redshift
($z \alt$3) extrapolation of the SFR. Therefore, for the neutrino
fluxes from PopII shown as comparison in the following, we have used
at low-redshift the GRB rate given in Eq. (12c) of
\cite{Murase:2005hy} normalized in order to match the observe GRB
rate $z$=0.

\begin{table}
\begin{center}
\begin{tabular}{|l|l|l|}
\hline Model & Reference & Notes\\
\hline
\hline CF0$5_a$ &  \cite{Choudhury:2004vs} &  high ion. efficiency\\
\hline CF0$5_b$ & \cite{Choudhury:2004vs} &  low ion. efficiency\\
\hline CF06 & \cite{Choudhury:06} & WMAP3, $S(M)$\\
 \hline
\end{tabular}
\caption{The models of reionization we consider for calculating the
neutrino fluxes. See text for details.\label{EoRModels}}
\end{center}
\end{table}

%%%%%%%%%%%%%%%%%%%%%%%%%%%%%
\subsection{Rate of GRBs from PopII}
%%%%%%%%%%%%%%%%%%%%%%%%%%%%%
If we denote by $\RII$ the SFR of PopII stars (taken from the models
reported in Table \ref{EoRModels}), an estimate for the number
density rate of GRBs from PopII stars, $\grbII$, can be obtained as
\be \grbII(z)= \sII\,\gII\,\frac{\RII(z)}{\MII}\, \label{grbII} \ee
where $\MII$ is the average mass of a PopII star, $\sII$  is the
fraction of PopII which are likely to form core collapse SNe and
$\gII$ is the fraction of core collapse SNe which are likely to form
a GRB. The average mass of a PopII star is obtained via an integral
over their IMF, $\imfII$, as
\be \label{PopIImass} \MII=\frac{\int \imfII(M)\,M\,\d M}{\int
\imfII(M)\,\d M}\,, \ee
assuming (consistently with all the EoR models considered) that
$\imfII(M)$ is a Salpeter mass function $S(M)$ which writes
\cite{Kroupa:2000iv}
\begin{eqnarray}
\label{eq:Salpeter} S(m) \propto\ m^{- \alpha_i}\left\{
\begin{array}{ll}
      \alpha_0 = 0.3\,, &  0.01<m< 0.08\\
      \alpha_1 = 1.3\,,   \ \
        &  0.08<m<0.50 \\
      \alpha_2 = 2.3\,,   \ \
        &  0.50<m\,, \\
\end{array}
\right.
\end{eqnarray}
where $m$ is the mass in solar units $M_{\odot}$. This is also used
to estimate $\sII$, assuming that all PopII with $m>10$ will end
their lives as SN,
\begin{equation}\label{eq:SNfrac}
\sII \simeq \frac {\int_{10}^{125} S(m) \d m}{\int_{0.1}^{125} S(m)
\d m}\,.
\end{equation}

We allow for a $z-$dependence in $\gII$ to take into account the
GRB--metallicity anti--correlation and  the fact that the
metallicity evolves with the age of the Universe. Following
\cite{Yuksel:2006qb} and  \cite{Yoon:2006fr},
\be \gII(z) =\gII(0)(1+z)^{1.4}=\frac{(1+z)^{1.4}}{1250}\,, \ee
where we fixed $\gII(0)$ according to \cite{Yoon:2006fr} and
$\gII(z)$ has been considered constant for $z > 4.5$, thus
considering no metallicity evolution above this redshift (however,
this choice affects very little the final results, most of the
contributions to the GRB neutrino background coming from bursts that
occur at $z \lesssim 7$).

%%%%%%%%%%%%%%%%%%%%%%%%%%%%%
\subsection{Rate of GRB from PopIII}
%%%%%%%%%%%%%%%%%%%%%%%%%%%%%
Formally, the number density rate of GRB from PopIII stars,
$\grbIII$ can be written identically to Eq. (\ref{grbII}),
namely
\be \grbIII(z)= \sIII\,\gIII\,\frac{\RIII(z)}{\MIII}\,, \ee
where the various coefficients and functions involved now assume
different values. The $\RIII(z)$ are taken from the models reported
in Table \ref{EoRModels}. Concerning the IMF of PopIII stars,
$\imfIII$, there are currently two hypotheses: $(i)$ a bimodal IMF
with a first peak at a few solar masses and the second one at $\sim
100\,M_{\odot}$ \cite{Nakamura:01}; $(ii)$ a peaked IMF around the
range of few hundred solar masses has been later presented in
\cite{Abel:2001pr}. Recently it has been pointed out,
\cite{O'Shea:2005si}, that primordial stars forming in a halo
neighboring the formation star of a PopIII have still zero
metallicity but smaller accretion rates onto the central dense
region, resulting in a final stellar mass of 10--100M$_{\odot}$.
This ``intermediate" generation would have equally contributed to
the reionization and we do implement this hypothesis by studying the
contribution of $M\sim 60\,M_{\odot}$ in our neutrino models.
Whatever the details are, the hypothesis $(ii)$ is nowadays widely
accepted in the community, or in any case it is the $\imfIII$ most
frequently implemented in EoR models. We shall consider the
following cases:
\begin{itemize}
\item[$(a)$]
$$\imfIII\propto \delta(\MIII-300\,M_{\odot})\,,$$
i.e. a toy model of a ``monochromatic" high mass mode for the IMF.
\item[$(b)$]
$$\imfIII\propto \delta(\MIII-60\,M_{\odot})\,,$$
i.e. a toy model of a monochromatic IMF, with the average mass of
metal--free stars forming from the collapse of halos triggered by
the explosion of the very first stars.
\item[$(c)$]
$$\imfIII\propto S(M)\,,$$
as assumed e.g. in \cite{Choudhury:06}.
\end{itemize}

Note that in both case $(a)$ and case $(b)$ one has $\sIII\simeq 1$,
while in case $(c)$ one obtains an expression analogous to Eq.
(\ref{eq:SNfrac}).

Concerning $\gIII$, there are no firm estimates. For the sake of
clarity and to simplify the comparison with previous results of
\cite{Schneider:2002sy}, we shall assume the same value,
\be
\gIII\simeq 1\,. \ee
This corresponds to an upper limit to the PopIII GRBs; a smaller GRB
efficiency would result in a down-scaling of the results presented
in the following.

%%%%%%%%%%%%%%%%%%%%%%%%%%%%%
\subsection{Neutrino yields from GRBs}
%%%%%%%%%%%%%%%%%%%%%%%%%%%%%
\begin{table}[!tb]
\begin{center}
\begin{tabular}{|c|c|c|c|c|c|}
\hline Model & Mass ($M_{\odot}$) & $\Eiso$ (erg) & $\Ejet$ (erg) & $\ris$ (cm)& $\eta$\\
\hline
\hline Hidden-A & 200 & $10^{54}$ & $10^{52}$ & $10^{13.0}$ & 1\\
\hline HIdden-B & 200 & $10^{54}$ & $10^{51}$ & $10^{13.0}$ & 1\\
\hline Hidden-C & 60 & $10^{53}$ & $10^{51}$ & $10^{12.5}$ & 1\\
\hline Hidden-D & 60 & $10^{53}$ & $10^{50}$ & $10^{12.5}$ & 1\\
\hline Hidden-E & 200 & $10^{54}$ & $10^{52}$ & $10^{11.0}$ & 10\\
\hline Hidden-F & 60 & $10^{53}$ & $10^{51}$ & $10^{11.5}$ & 10\\
\hline \hline $Prompt$ & 35-125 & $10^{53}$ & $1.24 \times 10^{51}$ & $10^{13-15.5}$ & 1\\
 \hline
\end{tabular}
\caption{The parameters varied in the models of GRB
considered.\label{popIIIModels}}
\end{center}
\end{table}
High energy neutrinos from GRBs are predicted in various scenarios.
The one most frequently discussed is neutrino emission in the
internal shocks that produce prompt emission \cite{Waxman:1997}. In
this scenario, accelerated protons interact with gamma-rays via
photomeson production, and produce pions and kaons, which decay into
neutrinos. The observed gamma-ray spectra are usually represented by
a broken power-law. As for proton spectra, a first order Fermi-type
spectrum is frequently assumed. This possibility has been studied by
many authors. Murase \& Nagataki \cite{Murase:2005hy,Murase:2006b}
did such calcultions using Geant4 with experimental data. They also
took into account various cooling processes. One parameter set of
their results (which is used in \cite{Achterberg:2007}) will be
shown for comparison in this paper, and named {\it Prompt} in the
following.

Another model was first suggested by M\'esz\'aros and Waxman
\cite{Meszaros:2001ms} and later extended in \cite{Razzaque:2003}.
The GRB progenitor is usually taken to be a massive star with a He
core and H envelope. The central engine of GRBs is still a major
open problem in astrophysics; the leading model for (long duration)
GRBs is a huge stellar collapse leading to formation of a highly
rotating black hole with an accretion disk (collapsar model). The
jet which produces succeeded GRBs or choked GRBs propagate in stars.
This jet would interact with the stellar envelope before it produces
prompt emission. In addition, the observed submillisecond
variability allows us to expect that intenal shocks can occur at
sufficiently small radii, where the Thomson optical depth exceeds
unity. Protons can be accelerated in such internal shocks, and they
can interact with photons from electrons that are accelerated in
internal shocks and/or termination shocks\footnote{More details about the terminology and 
GRB modeling are given in Appendix \ref{modelsummary}.}. 
Photons from such inner radii cannot escape due to large optical thickness, 
so that they are hidden sources as far as gamma-rays are concerned. Only neutrinos
would be useful as the probe of physical processes in this region.
Since here we want to estimate the high energy neutrinos from PopIII
GRBs, consistently with the empirical suggestion from Sec.
\ref{sec:Obser} we shall only evaluate neutrino emission from such
inner radii. Indeed, this emission is present also for chocked GRBs, and we 
stress that such ``hidden'' models are the ones that best match our choice
$\gIII \simeq$1. In fact, models such as the {\it Prompt} would predict a
lower ratio of GRBs, thus creating friction with our choice.
Throughout the paper, we take into account neutrino oscillations in
vacuum by adopting mixing angles
$\theta_{12}=0.59\,,\theta_{23}=\pi/4\,,$ and $\theta_{13}=0$. The
yields shown in the rest of this paper refer to the muon neutrino
flavor, which is the best channel for detection of a low-statistics
signal at neutrino telescopes given the long tracks of muons
produced in charged current interactions. For diffuse signals
competing with the atmospheric background, for which the separation
must be done on the basis of the different energy spectrum, the
electron neutrino showers may be an interesting observable
\cite{Beacom:2004jb}. However, showers are only detected if
contained in the instrumented volume, which lowers the effective
collecting area of the telescope with respect to the muon track
case. We shall see that our event rate predictions for muon tracks
fall below the sensitivity of a km$^3$ instrument, which makes
hopeless exploring alternative channels.

\begin{figure}[!htp]
\begin{tabular}{c}
\epsfig{file=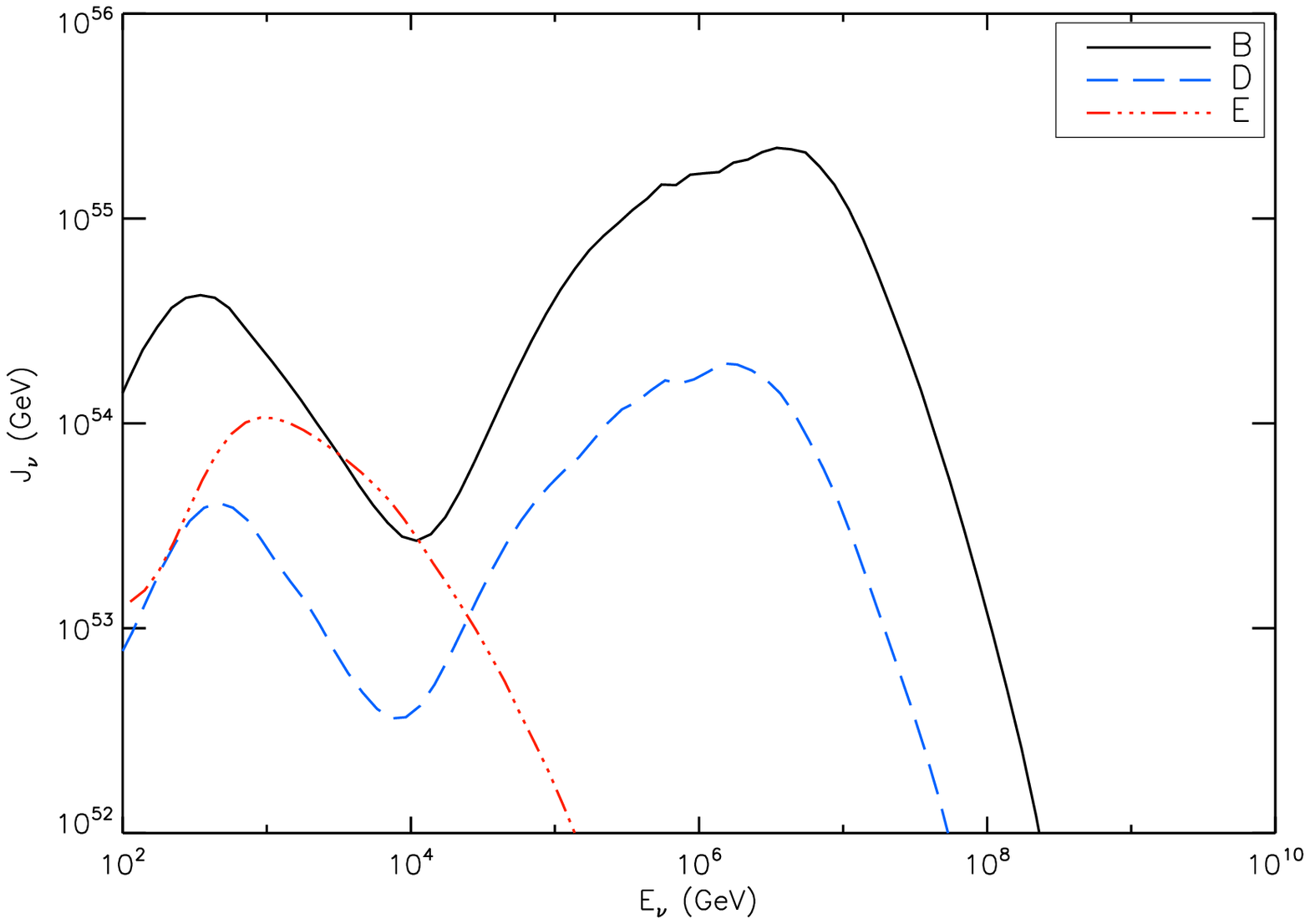,width=\columnwidth}  \\
\epsfig{file=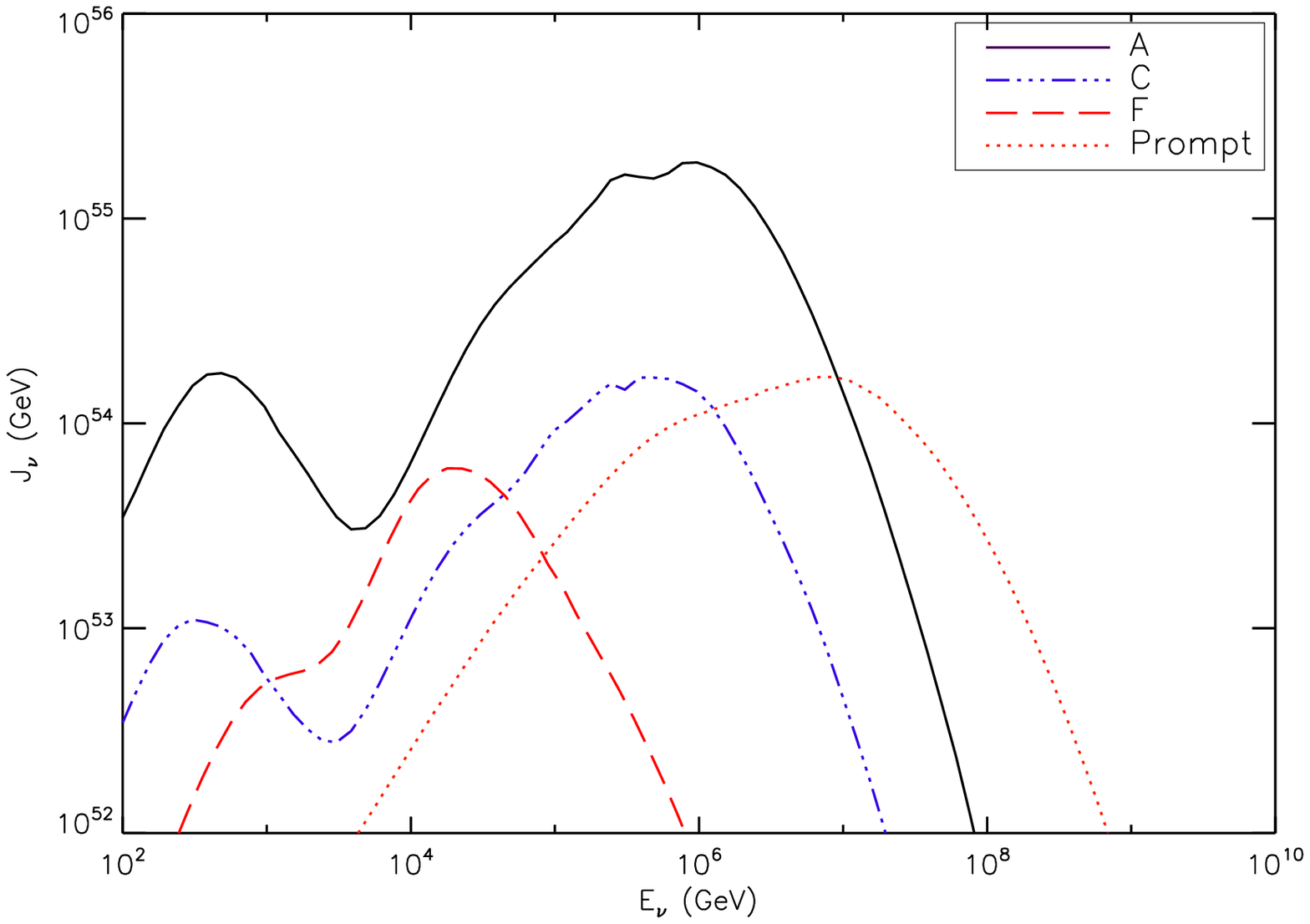,width=\columnwidth} \\
\end{tabular}
\caption{$J_{\nu}$ for the different models used, see Tab.
\ref{popIIIModels}.} \label{Ja}
\end{figure}

Since our knowledge of the GRB engine and PopIII properties is still
limited, it is unclear how efficient are PopIII stars as GRB
progenitors. But rotation and chemical mixing seem to be a key
ingredient towards successful collpsar models
\cite{Maeder:2000,Heger:2000}, and recent simulations in
\cite{Yoon:2006fr} suggest that GRBs can be born from PopIII stars,
an hypothesis we will assume throughout. We expect that more massive
star collapses lead to larger released energies \cite{Heger:03}.
Observationally, to ease the energy requirements, it is usually
considered that a globally asymmetric, relativistic jet is launched
from GRB progenitors. In fact, this picture is suggested by some
observations, e.g. \cite{Frail:2001}. The Ghirlanda
relation \cite{Ghirlanda:2004} implies that bursts can have various
jet energies, and some GRBs may have $\Ejet \sim
{10}^{52}$ erg, which is larger than the frequently used value
$\Ejet \sim 1.24 \times {10}^{51}$ erg. The isotropic energy is
highly uncertain, too, and it may be as high as $\Eiso \sim
{10}^{54}$ erg; we adopt this value also to allow direct comparison 
with previous results in \cite{Schneider:2002sy}.

Because of these high uncertainties, we explore a
wide range of parameter values for $\Ejet$ and $\Eiso$, as reported
in Table \ref{popIIIModels}. The other parameters varied are $\eta$,
a factor determining the maximum acceleration energy of protons, and
$\ris$.
Further details on the
model and its parameters are in Appendix \ref{modelsummary}.

The radial density profiles $\delta(r)$ for the PopIII progenitor
mass-models, $M=200M_{\odot}$ and $M=60M_{\odot}$, have been
calculated in \cite{Heger:2001cd} and \cite{Rockfeller:06},
respectively. We would like to comment on the apparent contradiction
between the $\imfIII\propto \delta(\MIII-300\,M_{\odot})$ motivated
in the previous section and the $M=200\,M_{\odot}$ model used for
the calculation for the neutrino flux. Physically, stars with masses
$M=140\,M_{\odot}<$M$<M=260\,M_{\odot}$ are known to directly
explode as PISNe without the formation of a Black Hole, thus being
inefficient GRB progenitors. While important for the GRB rate,
however, we expect that the differences in the density profile for
the two star masses are much less pronounced in the neutrino yields,
and certainly within the uncertainties and assumptions made in such
an uneven ground.

In Fig. \ref{Ja} we show the neutrino emission spectra at the
redshift of the source $J_{\nu,\rm III}(E)$ from the PopIII GRB
models considered in Table \ref{popIIIModels}, also compared with
the spectrum $J_{\nu,\rm II}(E)$ for the PopII model {\it Prompt}.

\begin{figure}[!t]
\begin{tabular}{c}
\epsfig{file=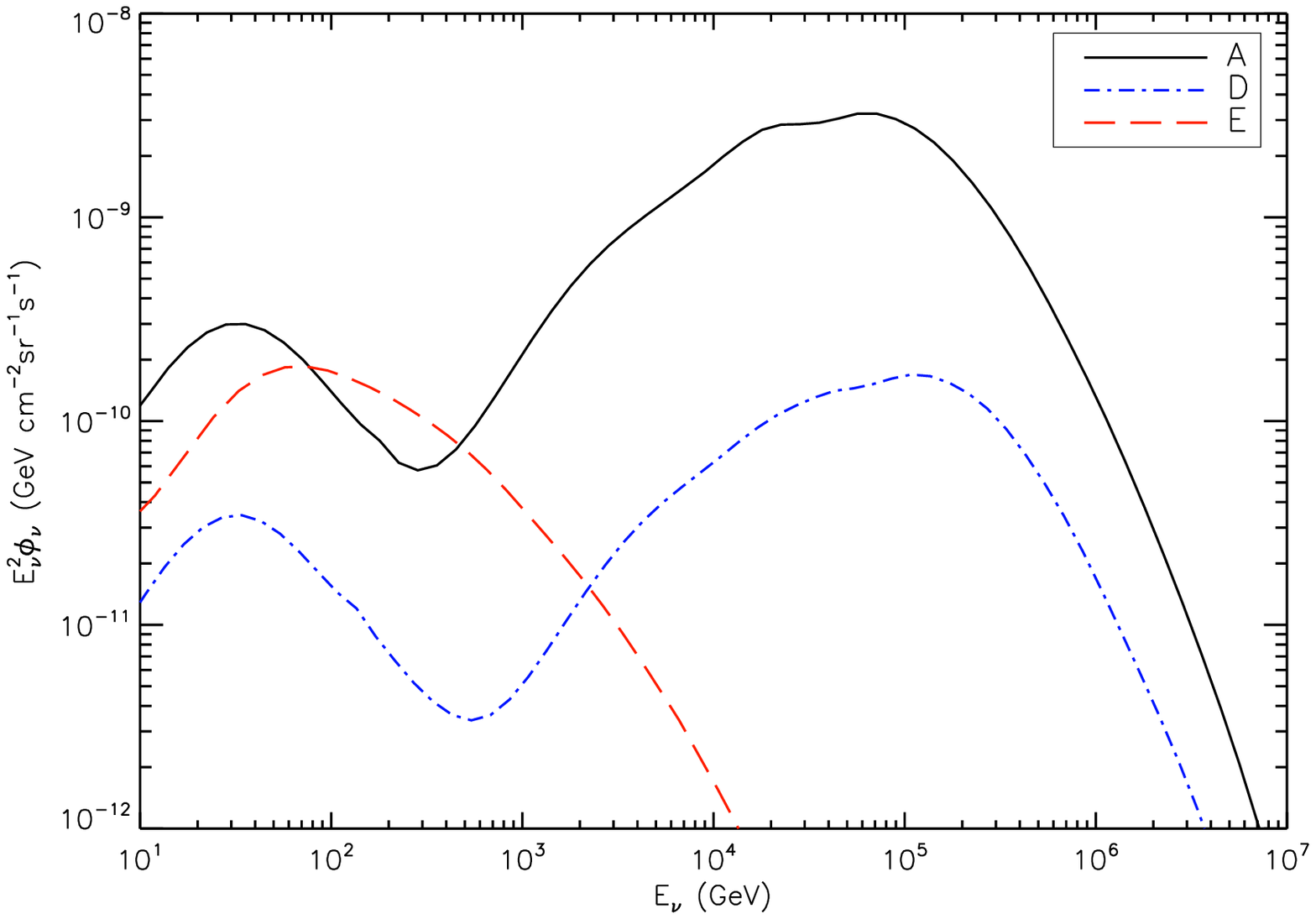,width=\columnwidth}  \\
\epsfig{file=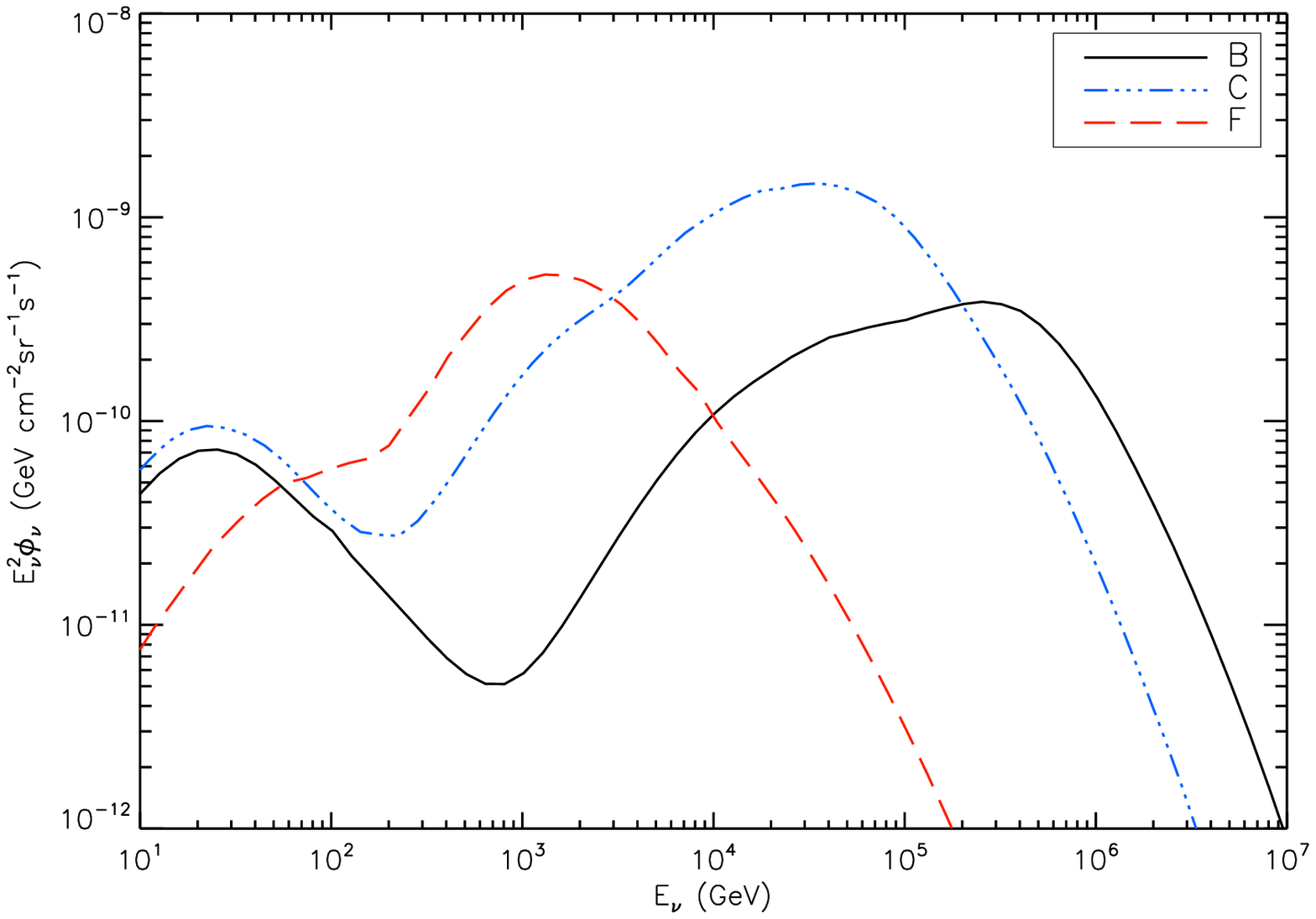,width=\columnwidth} \\
\end{tabular}
\caption{The $E_{\nu}^2\Phi_{\nu,\rm III}$ for different $J_{\nu,\rm
III}$ models, assuming CF05$_a$ as EoR model.} \label{Fa}
\end{figure}

In the scenario where internal shocks and termination shocks occur
at sufficiently small radii inside the progenitor star, almost
accelerated protons are depleted by hadronic and photomeson
reactions. However, produced mesons and muons suffer from cooling
processes. The strong magnetic field and copious photon field make
pions and muons cooler before they decay. Hence, the neutrino flux
will be suppressed due to cooling processes. In Fig. \ref{Fa}, the
relative flux level basically reflects the differences among
isotropic energies of models. However, the strong field makes mesons
and muons give up their energies, so that such suppression becomes
important. For example, the difference on the flux level between
Model E and Model F is reduced. Furthermore, the strong field 
can force the higher break energy, which is determined by equating the 
pion lifetime and its cooling time, to be shifted to lower energies. 
This effect also leads
to a double peak structure as shown in Fig.1 for Model A-D. The
lower peak corresponds to the contribution from the termination
shock. Here, in fact, nonthermal protons accelerated in the internal 
shocks interact with thermal photons produced by electrons 
accelerated in the termination shock.
On the other hand, the higher peak id due to protons from the
internal shocks themselves. If we assume similar equipartion
parameters, the termination shock can lead to the stronger field
than that of internal shocks. Therefore, neutrinos from this region
have lower energies due to pion and muon cooling and this explains the double
peak structure. However, the prominence of this feature
is probably due to the simplicity of the model considered: in
realistic situations, the emission regions would not be approximated
by the simplest two zone model (internal shock region and
termination shock region), and the gradient of the field strength
should be taken into account. Then, the double peak structure would
be more smoothed, although a detailed treatment of this issue
goes beyond the scope of this paper.

%%%%%%%%%%%%%%%%%%%%%%%%%%%%%%%%%%%%%%%%%
\section{Results and discussion}\label{sec:PrelRes}
%%%%%%%%%%%%%%%%%%%%%%%%%%%%%%%%%%%%%%%%%
In this section we present our results for the neutrino fluxes
expected at the Earth. In Fig. \ref{Fa} we show the contribution of
PopIII stars calculated as described in the previous sections,
assuming CF05$_a$ as a fiducial EoR model. In general, these fluxes
present a huge variability due to the large uncertainties on the
production parameters, yet they are all much lower than the most
optimistic results found in \cite{Schneider:2002sy}.

\begin{figure}[!htbp]
\begin{center}
\epsfig{file=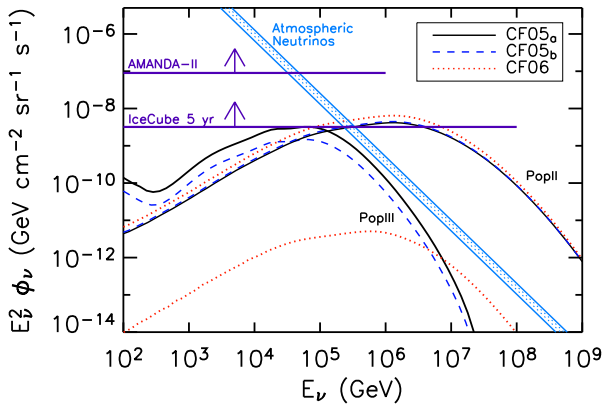,width=\columnwidth} \caption{The muon neutrino
flux $E_{\nu}^2\Phi_{\nu}$ for the different Reionization models,
using the  A model for $J_{\nu,\rm III}$. Upper-right set of curves
refer to the PopII flux, lower-left ones to the PopIII contribution
under optimistic assumptions (see text).}
\label{plotresults}\label{Fb}
\end{center}
\end{figure}

In Fig. \ref{Fb} we compare the high energy neutrino flux at Earth
expected from PopII stars with the one from PopIII for different but
mutually consistent EoR models, as denoted by the same line-style.
When accounting for the assumed baryon loading factor of $\xi_{\rm
acc}=10$ (see Appendix for the role of this parameter) the estimated
contribution from PopII stars agrees with typical results found in
the literature, and the dependence on the EoR model used is
marginal, reflecting the fact that at low redshifts (where most of
the signal comes from) all models must agree with the available data
\footnote{The slight discrepancy with previous results in the PopII
neutrino flux at Earth is due to the higher GRB rate we use for
$z>3$, arising from the reionization models used here.}. On the
other hand, the situation for PopIII is different. The solid and
dashed lines assume the model A neutrino yields, which as shown in
Fig. \ref{Fa} maximizes the neutrino production. The dotted line
instead follows from the CF06 model, where we have implemented a
$Prompt$ spectrum for $J_{\nu,\rm III}$ coherently with the
assumption of a Salpeter IMF. We have treated this case as the PopII
one, thus obtaining the fraction of collapsars $\gIII\simeq
10^{-2}$, consistently with no metallicity evolution beyond $z=
4.5$. This illustrates the high energy neutrino flux for the case of
an hypothetical low--mass PopIII generation; whereas the CF06 SFR
coupled with the A model for $J_{\nu,\rm III}$ gives results almost
indistinguishable from CF05$_b$ and we have therefore not plotted
it. The solid and dashed lines show that the SFR has only a minor
effect on the neutrino flux at Earth; on the other hand, their
comparison with the dotted line shows that the IMF plays a
fundamental role in shaping the flux at Earth, affecting the
fraction of stars that will give rise to a collapsar and the
magnitude of their explosion energy.

A fairly robust outcome of our study is however that neutrinos from
PopIII GRBs will not be detectable with current or near future
experiments. In fact, the PopIII flux shown in Fig. \ref{Fb} falls
below both the current AMANDA-II bound (4-years data, see
\cite{Halzen:2006mp}) and future prospects for five years of IceCube
exposure. Even worse, this contribution is buried beneath the
atmospheric neutrinos, whose average spectrum and uncertainty is
plotted in Fig \ref{Fa} according to the compilation recently
reported in \cite{Evoli:2007iy}. Also, it is worth stressing that
while the PopII flux we have plotted in Fig \ref{Fb} is a relatively
robust expectation because the background can be evaluated from the
observed GRB rate, the PopIII one is an upper limit, as it assumes
that all the stellar events will give rise to a collapsar with high
energy neutrino emission. These results greatly disagree with the
much more optimistic estimates obtained in \cite{Schneider:2002sy},
thus motivating some explanation. We believe that our discrepancy
with previous results is to impute to the different normalization of
PopIII SFR. In particular, from Fig. 1 in \cite{Schneider:2002sy}
one can infer that the neutrino spectra results plotted in their
Fig. 2 require a rate of $\sim10^9$ GRBs per year versus e.g. a rate
of $\sim 10^5$ GRBs per year for the CF05$_a$ model adopted here
(more details can be found in \cite{Iocco:2007qy}). Actually, such a
large discrepancy only arises for the most extreme model considered
in \cite{Schneider:2002sy}, which adopts a value for the ionizing
efficiency which is far too large, based on the results of
self-consistent modeling of the reionization epoch nowadays
available. Predictions for more realistic models are in better
agreement with our results. It is also interesting to note that the
theoretical models used here are consistent, within a factor of a
few, with the estimate for the overall GRB rate obtained empirically
in Sec. \ref{sec:Obser}, which makes us confident on the more
realistic estimates provided here with respect to previous
literature.

%%%%%%%%%%%%%%%%%%%%%%%%%%%%%%%%%%%%%%%%%
\section{Conclusion}\label{conclusion}
%%%%%%%%%%%%%%%%%%%%%%%%%%%%%%%%%%%%%%%%%
We have performed a study of the high energy neutrino diffuse
background which is to be expected from Population III stars, under
the assumption that they will end their lives as GRB. We have
compared it with the analogue expected from Population II stars,
using mutually consistent SFRs obtained from EoR models available in
literature. Our estimate of the PopII GRB rate has been performed
under widely accepted assumptions and leads to estimates of the high
energy neutrino fluxes which are in agreement with previous results
\cite{Murase:2005hy,Razzaque:2003}. On the other hand, we have
presented a maximal model for PopIII stars, assuming that all of
them will end their lives with a GRB, either choked or not. Even
under such optimistic assumptions, a detection of the diffuse high
energy neutrino background expected from Population III stars
appears out of reach. Even worse, the contribution from PopII GRBs
would contaminate that from PopIII GRBs. In addition, we cannot
expect neutrino signals correlated with gamma-rays from GRBs,
because it is thought to be very rare to see PopIII GRBs by the
current satellite such as Swift. Hence, the PopIII neutrino signals
are expected to be hidden by atmospheric neutrino background. With
reasonable values of the nonthermal baryon loading factor, an
extreme contribution from PopIII GRBs---including possible choked
bursts---falls in fact underneath IceCube five years sensitivity and
is overwhelmed, at low energies, by the atmospheric neutrino
background. This implies that although PopIII stars may contribute
to the high-energy diffuse neutrino background and their spectrum
would be indeed sensitive to their IMF and SFR, neutrinos cannot be
used as a diagnostic tool to check properties of either the
population of stars during the epoch of Reionization or the GRB
internal shock properties, thus confirming the elusive nature of the
earliest generation of stars.

\section*{Acknowledgments}
F.I. thanks A. Heger and C.L. Fryer for kindly providing the stellar
models, and A. Ferrara for kind clarifications about EoR models.
P.S.~acknowledges support by the US Department of Energy and by NASA
grant NAG5-10842. This work is in part supported by a Grant-in-Aid
for the 21st Century COE ``Center for Diversity and Universality in
Physics'' from the Ministry of Education, Culture, Sports, Science
and Technology of Japan. S.N. is partially supported by
Grants-in-Aid for Scientific Research from the Ministry of
Education, Culture, Sports, Science and Technology of Japan through
No. 19104006 and 19740139. K.M. is partially supported by Japan
Society for the Promotion of Science.

\appendix
%%%%%%%%%%%%%%%%%%%%%%%%%%%%%%%%%%%%%%%%%
\section{Modeling neutrino emission from Jets inside the GRB Progenitor Star}\label{modelsummary}
%%%%%%%%%%%%%%%%%%%%%%%%%%%%%%%%%%%%%%%%%
Here we summarize the basic mechanisms by which neutrinos are produced in
jets inside the GRB progenitor star.

We assume that a (long) GRB is a huge stellar collapse leading to
the formation of a highly rotating black hole with an accretion
disk, during which a globally asymmetric, relativistic jet is
launched from the GRB progenitor. The GRB observations imply that
the Lorentz factor of jets is very large ($\Gjet \sim 100$) in the
gamma-ray emitting region. This is the final value which is achieved
outside the star, and it may be representative of the intrinsic
injection Lorentz factor. Of course, we do not know the intrinsic
value and it may also be much smaller than the final value. We shall
briefly comment on the latter case later. When the jet propagates
inside the star, it will produce a bow shock ahead of it. This jet
is capped by a termination shock and a reverse shock. The jet
termination radius (where the jet is decelerated) is indicated by
$\rh$. If the jet is highly variable and the variability time scale
$\delta t$ is small enough, internal shocks can also occur inside
the pre-decelerated jet. The internal shock radius is written as
$\ris\approx 2 \Gjet^2\, c\, \delta t$. In the usual internal shock
scenario of GRBs, the observed gamma-rays are attributed to internal
shocks which occur in the optically thin region outside the
progenitor star. Lower $\Gjet$ and/or $\delta t$ lead to smaller
radii below the stellar surface, which will be optically thick.
Here, we are interested in such opaque, sub-surface internal shocks,
and we assume $r_{\rm{is}} \lesssim r_{h} < r_{*}$, where $r_{*}$ is
the radius of the progenitor's stellar surface.

These internal shocks are expected to be collisionless, so that both
electrons and protons may be accelerated. If the magnetic field is
strong enough, electrons can be accelerated up to high energies and
radiate photons. The photon density in the pre-decelerated jet is
given by
\begin{equation}
\Ug = \frac{E_{\rm{sh}}}{4 \pi r_{\rm{is}}^2 \Gjet\, \Delta l}\,,
\end{equation}
where $\Delta l$ is the width of subshells, for which we use $\Delta
l \approx r/\Gjet$ because we have assumed that the jet acceleration has already
ceased. The magnetic energy density in the field $B$ is expressed as
\begin{equation}
\UBjet=\frac{B^2}{8\pi}=\xBjet\,\Ug\,.
\end{equation}

Due to the existence of the strong magnetic field, electrons radiate
synchrotron photons. These photons will be thermalized, if the jet
is optically thick to Thomson scattering. We assume $\xBjet=0.1$ and
expect that the jet is opaque. Therefore, we can approximate the
spectral distribution of the radiation of energy density $\Ug$ by a
black-body spectrum, and associate to it a temperature $\Tjet$. The
above approximate treatment is accurate enough for our purpose.

Protons get accelerated in the shocks as well. It is widely believed that
cosmic rays can be accelerated by the first-order Fermi acceleration mechanism.
We assume that this mechanism can work efficiently and adopt the spectral index $\sim 2$.
The non-thermal proton
energy density is expressed as $\Up = \xacc \Ug$, where $\xacc$ is
the nonthermal baryon loading factor and for efficient proton acceleration we can express
 as $\xi_{\rm{acc}} \sim 1/\epsilon_e$, $\epsilon_e$ being the fraction of
internal energy carried by the electrons.
Because plausible values of this parameter is
not known yet, we adopt $\xacc=10$, which corresponds to the
assumption that the energy of protons per logarithmic energy bin is
comparable to the GRB radiation energy
\cite{Waxman:1997,Murase:2005hy}. Too large value of $\xi_{\rm acc}$
requires too small values of $\epsilon_e$, which is usually
unexpected in GRBs (but there is no proof because we do not know the
total explosion energy). The non-thermal proton spectrum is given by
\begin{equation}
 \frac{d \npj}{d \varepsilon _p} = \frac{\Up}{\ln
(\varepsilon _p^{\rm{max}}/\varepsilon_p^{\rm{min}})}
\varepsilon_p^{-2}
\end{equation}
The minimum energy $\varepsilon _p^{\rm{min}}$ is set to $10$ GeV.
The choice of this value is not so sensitive to our final results.
On the other hand, the maximum energy is important for the purpose
of knowing neutrino spectra at the highest energies. The maximum
energy is determined by the condition
\begin{equation}
\frac{e\,B c}{\eta \varepsilon _p^{\rm{max}}}=\tacc^{-1} \simeq
\sum_i t_{i}^{-1}
\end{equation}
where at the l.h.s. it appears the Larmor radius of the proton times
the pre-factor $\eta (={\cal O}(1-10))$ which depends on the details
of acceleration mechanism; $\tacc$ is essentially
the acceleration time in the Bohm limit, and the
sum at the r.h.s. extends over all the energy loss channels
timescales $t_i$.

Accelerated protons can interact with protons themselves or photons,
which lead to meson production such as pions and kaons. In this
paper, we consider neutrinos produced through photomeson production
only. The method of calculation is described in
\cite{Murase:2005hy,Murase:2007}. Here, we briefly sketch this
process by simple analytic considerations. The photomeson production
is a threshold process, with a threshold energy of about $145$ MeV
in the rest frame of the incident proton. The dominant inelastic
channel is $p \gamma \rightarrow {\Delta}^{+}$ with the cross
section $\sigma_{p \gamma} \approx 5 \times {10}^{-28} \,
{\rm{cm}}^2$ and the inelasticity $\kappa_p \approx 0.2$, which is
called $\Delta$-resonance. For the cases we consider, the $p \gamma$
optical depth at the $\Delta$-resonance is very large, so that the
photomeson production efficiency can also be high. This means that
almost all the protons that have sufficiently high energies (above
the threshold energy) will be depleted due to photomeson production.

The threshold for $pp$ inelastic interactions is lower, so these
processes will also occur.  Although this may be an important
neutrino source in the TeV energies (in particular for low Lorentz
factors $\Gjet$), in this paper we focus on high energy neutrinos
produced by sufficiently high energy protons above the threshold
energy for photomeson production. We take into account the $pp$
process only for estimating the proton energy loss time scale,
treating this process analytically for simplicity.

Next, we consider the interaction between the jet and the progenitor
star. As the jet advances through the star, it drives a bow shock
ahead of it. The jet is capped by a forward shock, and a reverse
shock moves back into the jet, where the relativistic jet is
decelerated. The shocked jet plasma and shocked stellar plasma would
advance together with a jet head Lorentz factor $\Gh \ll \Gjet$. By
equating the pressures behind the forward shock and reverse shocks,
one finds the following estimate for the Lorentz factor of the jet
head,
\begin{equation}
\Gh \simeq \Gjet^{1/2} {\left( \frac{m_p\, \npj}{4\, \delta}
\right)}^{1/4}\,,
\end{equation}
where $m_p$ is the proton mass and $\delta$ is the mass density of
the environment.

The relative Lorentz factor between the shocked jet plasma and
un-shocked jet plasma is
\begin{equation}
\Gprime \simeq \frac{1}{2} \left( \frac{\Gjet}{\Gh} +
\frac{\Gh}{\Gjet} \right)\,.
\end{equation}

Electrons will be accelerated in these shocks. However, the
electrons would give up all their energy on a very short time scale,
by synchrotron and inverse-compton (IC) cooling, converting a large
fraction of the shocked plasma internal energy into radiation. These
radiated photons in the shocked jet plasma will be thermalized due
to large optical thickness. Hence, target photon density will be
approximately a black-body radiation, with an overall energy density
\begin{equation}
\Ugh \simeq (4\Gamma^{\prime}+3)(\Gamma^{\prime}-1)\npj\, m_p\,
c^2\,,
\end{equation}
with an associated temperature $\Th$.

The reverse shock is likely to become radiation dominated, so that
the IC cooling by the electrons becomes important and affects the
dissipation of jet kinetic energy. If the reverse shock is indeed
radiation dominated, the shock thickness would also be of order the
mean-free path of thermal photons propagating into the jet. The
copious photon field also affects the neutrino spectrum. Charged
mesons produced via $p\gamma$ (and $pp$) interactions will suffer IC
and synchrotron losses, and neutrino spectra will be suppressed. The
magnetic energy density in the shocked jet frame is expressed as
\begin{equation}
\UBh = \xBjh\,\Ugh\,.
\end{equation}
For $\xBjh={\cal O}(0.1-1)$, the magnetic field strength is very
large, up to $\sim {10}^{7-9}$ G. Hence, charged particles would
suffer from synchroron loss, and one should take into account
synchrotron losses of pions in order to calculate neutrino spectra,
as we do in our computation.

If non-thermal protons that are accelerated in internal shocks are
not completely depleted, they can enter the shocked jet.
Such protons are expected to interact with photons in the shocked jet region,
and to produce neutrinos.
In principle, protons can be accelerated at the termination shocks,
too. However, cooling processes such as IC loss significantly will reduce
the maximum proton energy there. Hence, we treat the contribution
from protons that are accelerated in internal shocks only. The
proton spectrum in the frame of the shocked jet plasma is given by
\begin{equation}
 \frac{d \nph}{d \varepsilon_p} \simeq (1- \fpj)\,
{\Gprime}^2\,\frac{d \npj}{d \varepsilon_p}\,,
\end{equation}
where $\fpj$ is a fraction of depleted protons in the internal
shocks, which can be estimated as the ratio of the dynamical
timescale involved with the typical energy loss time. As protons
approach the reverse shock, they can interact with photons via
photomeson production, if proton energy is above the threshold for
photomeson production. It turns out that also in this region the
photon density is high enough to ensure that almost all protons that
have sufficiently high energies are depleted due to photomeson
production. However, neutrino energy from this region is smaller
than that from the pre-decelerated jet.

The $pp$ interaction also occurs in the shocked jet plasma, and a
further neutrino source is due to non-thermal proton interactions
with cold protons in the star after they escape the shocked jet
plasma. For more detailed predictions, these processes should be
taken into account. Yet, given the high efficiencies of the
photohadronic processes for our typical parameters, we expect them
to be at most subleading at the high energies we are
mostly interested in.

Finally, note that synchrotron, IC, and adiabatic cooling of pions
and muons are important in shaping our final result. The treatment
of these cooling processes is also similar to that used in
\cite{Murase:2005hy,Murase:2006b}. Analytic considerations on
effects of these cooling processes can be found in
\cite{Razzaque:2003}.

\end{document}